\documentclass[aps,preprint,amsmath,amssymb]{revtex4}
\usepackage{graphicx}
\begin{document}
\title{Anomalous $tq\gamma$ couplings in $\gamma p$ collision at the LHC}

\author{M. K\"{o}ksal}
\email[]{mkoksal@cumhuriyet.edu.tr} \affiliation{Department of
Physics, Cumhuriyet University, 58140, Sivas, Turkey}

\author{S. C. \.{I}nan}
\email[]{sceminan@cumhuriyet.edu.tr}
\affiliation{Department of Physics, Cumhuriyet University,
58140, Sivas, Turkey}

\begin{abstract}
We have examined the constraints on the anomalous $tq\gamma$ ($q=u,c$) couplings through the process $pp\to p\gamma p\to pWbX$ at the LHC  by considering four forward detector acceptances: $0.0015< \xi < 0.5$, $0.0015< \xi <0.15$, $0.015< \xi < 0.15$ and $0.1< \xi < 0.5$ where $\xi = E_{\gamma}/E$
with $E_{\gamma}$ and $E$ the energies of the photon and of the incoming proton,
respectively. The sensitivity bounds on the anomalous couplings have been obtained at the $95\%$ confidence level in a model independent effective lagrangian approach. We have found that the bounds on these couplings can be highly improved compared to current experimental bounds.
\end{abstract}

\maketitle

\section{Introduction}
The top quark is the heaviest particle of the Standard Model (SM). Therefore, the top quark
properties, and their production process provide a possibility for
probing new physics beyond the SM. Furthermore, the impacts of new physics on the top quark couplings are considered to be larger than that on any other fermions, and conflicts with the SM expectations could be measured as described in \cite{agu}.  A search for rare decays of the top quark is one of such studies. The search for the top quark anomalous interactions via
Flavour Changing Neutral Currents (FCNC) is of special interest. For the top quark, FCNC
decays $t\rightarrow q\gamma$ ($q=u,c$) can not be seen at the tree level of the SM. These
decays can only make loop contributions. As a result, the branching ratios of $t\rightarrow q\gamma$ are very small, and
they are at the order of $10^{-10}$ \cite{he,ee1,ee2}. However, various extensions of the SM, such as the quark-singlet model \cite{8,9,10}, the two-Higgs doublet model \cite{11,12,13,14,15,16}, the minimal supersymmetric model \cite{17,18,19,20,21,22,23}, supersymmetry
 \cite{24}, the top-color-assisted technicolor model \cite{25}
or extra dimension model \cite{26,27} could lead to a huge enrichment of those kind of decays.

The CDF collaboration bounds on the branching ratios at $95\%$ C. L. for the process $t\rightarrow q\gamma$ as follows \cite{cdf}

\begin{eqnarray}
BR(t \rightarrow u
\gamma)+BR(t \rightarrow c
\gamma)<3.2\%.
\end{eqnarray}
Furthermore, the ZEUS collaboration obtained upper limits at $95\%$ C.L. on the
anomalous $tq\gamma$ couplings $\kappa_{tu\gamma}<0.12$ \cite{zeus}. The Large Hadron Collider (LHC) can produce a large number of top quarks. Therefore, top quark interactions can be examined with high sensitivity. In particular, ATLAS collaboration has predicted a sensitivity
of $BR(t \rightarrow q\gamma)\sim 10^{-4}$ at $5\sigma$ level \cite{he9}.

The FCNC effective Lagrangian among the top quark, two quarks $u$, $c$ and
the photon $\gamma$ can be written as \cite{zeus}

\begin{eqnarray}
\textit{L}= \sum_{q_{i}=u,c} g_e e_{t}\bar{t}\frac{i\sigma_{\mu\nu}p^{\nu}}{\Lambda}\kappa_{tq_{i}\gamma}q_{_{i}}A^{\mu}.
\label{lag}
\end{eqnarray}
Here $\kappa_{tq_{i}\gamma}$ is the anomalous coupling for the neutral currents with a photon;
$\Lambda$ is a new physics scale; $\sigma_{\mu\nu}=[\gamma_{\mu},\gamma_{\nu}]/2$; $g_e$ is the electromagnetic
coupling constant; $e_{t}$ is the electric charge of the top quark. $\Lambda$ is the conventionally taken mass of the top quark ($m_t$) for the sake of definiteness. Hence, we take $\Lambda=m_t$. Also, we assume in our calculations $\kappa_{tu\gamma}=\kappa_{tc\gamma}$. Using the anomalous interaction given in Eq.\ref{lag}, the decay width can be obtained as follows,

\begin{eqnarray}
\Gamma(t \rightarrow
q\gamma)=\frac{g_{e}^{2}e_{t}^{2}\kappa_{tq\gamma}^{2}m_{t}^{3}}{8 \pi \Lambda^{2}}         &&           (q=u,c)
\end{eqnarray}
where we put the masses of the u and c quarks equal to the zero.
Branching ratio of the anomalous $t\rightarrow q\gamma$ decay
can be given by the following equation, since the main decay mode of the top quark is $t\rightarrow b W$

\begin{eqnarray}
BR(t\rightarrow q\gamma)=\frac{\Gamma(t\rightarrow
q\gamma)}{\Gamma (t\rightarrow b W)}.
\end{eqnarray}
Using this equation, from the experimental constraints of the CDF collaboration it is easy to obtain magnitude of the upper limit on $\kappa_{tq\gamma}=0.29$.

\begin{figure}
\includegraphics{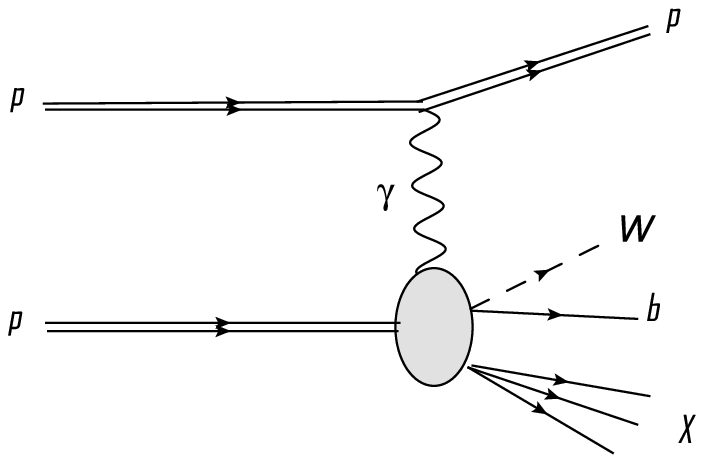}
\caption{Schematic diagram for the reaction $pp\to p \gamma p\to pWbX$.
\label{fig1}}
\end{figure}

In this work, we have examined anomalous FCNC interactions for the process $p p\to p\gamma p\to pbWX$ at the LHC. We show a schematic diagram for the this reaction in Fig.\ref{fig1}. The subprocess of the main reaction is $\gamma q\to Wb$. This process is becoming interesting as an additional way to investigate for SM or new physics.

In many  situations, ultraperipheral collisions and elastic interactions can not be detected at the central detectors.  Forward detectors are developed by the ATLAS and CMS collaborations to detect the scattering particles which can not be caught by the central detectors with limited pseudorapidity. These extra detectors are placed at distance of $220$ m - $420$ m from the central detectors. Usual $pp$ deep inelastic scattering (DIS) incoming protons dissociate into partons. Therefore, DIS interactions have very sophisticated backgrounds. In the DIS process, made up of jets from the proton remnants, some ambiguities are created which make it hard to detect the new physics signals beyond the SM. However, $\gamma\gamma$ or $\gamma p$ interactions have a clean environment compared to the usual proton-proton DIS, since in $\gamma\gamma$ or $\gamma p$ collisions with almost real photons, a photon is emitted, while the proton remains intact. Because of both of the incoming protons remaining intact, $\gamma\gamma$ collisions provide fewer backgrounds compared to the other processes. However, $\gamma p$ collisions have higher energy and effective luminosity with respect to $\gamma\gamma$ interactions.

In $\gamma p$ collisions, the almost real photons with low virtuality are emitted from only one of the proton beams and it is a good approximation to assume that they are on-mass-shell. Because of the low virtuality of the photons, the structure of the photon emitting protons are not spoilt. Also, almost real photons are scattered with small angles, and then they have a low transverse momentum. Since these photons have very high energy, they can interact with quarks in the other incoming proton's internal structure. On the other hand, intact protons which are emitting photons deflect slightly their path along the beam pipe, and, generally, they can not be detected in central detectors. One of the main properties of forward detectors is to detect the intact protons with some momentum fraction loss given the formula, $\xi=(|\vec{p}|-|\vec{p}^{\,\,\prime}|)/|\vec{p}|$, where $\vec{p}$ and $\vec{p}^{\,\,\prime}$ are momentums of incoming protons and intact scattered protons, respectively. At
very high energies, it is a good approximation to write $\xi=E_{\gamma}/E$ where $E, E_{\gamma}$ are the energies
of the proton emitting the photon and of the photon, respectively. If the forward detectors are established closer to central detectors, a higher $\xi$ can be obtained. Forward detectors can detect intact outgoing protons in the interval $\xi_{min}<\xi<\xi_{max}$. This interval is known as the acceptance of the forward detectors. ATLAS forward detectors have an acceptance of $0.0015 < \xi <0.15$ \cite{albrow} and CMS-TOTEM forward detectors are placed closer to the central detectors and the acceptances span $0.0015 < \xi< 0.5$, $0.1< \xi <0.5$ \cite{lhc6, avati}.

Photon-induced reactions in hadron collider phenomena were recently observed in the measurements of the CDF collaboration \cite{cdf1,cdf2,cdf3,cdf4,cdf5,cdf6,cdf7}, and these measurements are consistent in both theoretical expectations with $p\bar{p} \to p\ell^{+}\ell^{-}\bar{p}$ through two photon exchange ($\gamma\gamma \to \ell^{+}\ell^{-}$). Therefore, the photon-induced interactions' potential at the LHC is significant, with its high energetic pp collisions, and high luminosity \cite{albrow,lhc6,avati,lhc1,lhc1a,lhc2a,lhc2,lhc4,lhc5,lhc7,khoze,albrow2,inanc,inan,kepka,bil,bil2,kok,inan2,gru,inanc2,epl,inanc3,bil4}. Moreover, two photon reactions $pp \to p\gamma \gamma p \to p\mu^+ \mu^- p $, $pp \to p\gamma \gamma p \to p e^+ e^- p $, and $pp \to p\gamma \gamma p \to p W^+ W^- p $ have been measured by the CMS collaboration from the early LHC data at $\sqrt{s}=7$ TeV \cite{ch1,ch2,cmss}.

The photon-induced reactions in $pp$ collisions can be obtained in the framework of the equivalent photon approximation (EPA) \cite{budnev,baur}. In this approximation the equivalent photon spectrum, given the virtuality $Q^2$ and the energy of the quasi-real photons $E_\gamma$ ($E_\gamma>>Q^2$), is given as follows:

\begin{eqnarray}
\frac{dN}{dE_{\gamma}dQ^{2}}=\frac{\alpha}{\pi}\frac{1}{E_{\gamma}Q^{2}}
[(1-\frac{E_{\gamma}}{E})
(1-\frac{Q^{2}_{min}}{Q^{2}})F_{E}+\frac{E^{2}_{\gamma}}{2E^{2}}F_{M}]
\label{phs}
\end{eqnarray}

\noindent where $E$ is the incoming proton energy ($E_\gamma=\xi E$).
The remaining terms are as follows,

\begin{eqnarray}
Q^{2}_{min}=\frac{m^{2}_{p}E^{2}_{\gamma}}{E(E-E_{\gamma})},
\;\;\;\; F_{E}=\frac{4m^{2}_{p}G^{2}_{E}+Q^{2}G^{2}_{M}}
{4m^{2}_{p}+Q^{2}} \\
G^{2}_{E}=\frac{G^{2}_{M}}{\mu^{2}_{p}}=(1+\frac{Q^{2}}{Q^{2}_{0}})^{-4},
\;\;\; F_{M}=G^{2}_{M}, \;\;\; Q^{2}_{0}=0.71 \mbox{GeV}^{2}.
\end{eqnarray}

\noindent Here, $m_{p}$ is the mass of the proton, $\mu_{p}^{2}=7.78$ is the squared magnetic moment of the proton, $F_{E}$ and $F_{M}$ are functions of the electric and magnetic form factors, respectively, and $E, E_{\gamma}$ are the energies of the proton emitting the photon and of the photon, respectively. The cross section  for the main process $pp \to p \gamma p \to p W b X $ can be found by integrating $\gamma q \to W b$ subprocess cross section over the photon and quark spectra:

\begin{eqnarray}
\label{efflum}
\sigma(pp \to p\gamma p \to p W b X)&&=\sum_{q=u,c}\int_{Q^{2}_{min}}^{Q^{2}_{max}}
{dQ^{2}}\int_{\xi_{min}}^{\xi_{max}}{d\xi} \nonumber \\
&& \times \int_{x_{min}}^{x_{max}} {dx} \left(\frac{dN_{\gamma}}{d\xi dQ^2}\right)\left(\frac{dN_q}{dx}\right)\hat{\sigma}_{\gamma q \to Wb}(\hat{s}).\nonumber \\
\end{eqnarray}

\noindent where $x$ is the momentum fraction of the proton's momentum carried by the quark. $\frac{dN_q}{dx}$ is the quark distribution function of the proton. Also, we have taken the $Q_{max}^2=2$ GeV$^2$ since $Q_{max}^2$ greater than $2$ GeV$^2$ does not make a significant contribution to this integral. From Eq.\ref{efflum} the following equation can be obtained,

\begin{eqnarray}
\sigma(pp \to p\gamma p \to p W b X)&&=\sum_{q=u,c}\int_{Q^{2}_{min}}^{Q^{2}_{max}}
{dQ^{2}}\int_{\frac{M_{inv}}{\sqrt{s}}}^{\sqrt{\xi_{max}}}{dz} 2z \nonumber \\
&& \times \int_{MAX(z^2, \xi_{min})}^{\xi_{max}}\frac{d\xi}{\xi}\frac{dN_\gamma}{d\xi dQ^2}N_q\left(\frac{z^2}{\xi}\right)\hat{\sigma}_{\gamma q \to Wb}(\hat{s})\nonumber \\
\end{eqnarray}
where $M_{inv}$ is total mass of the final state particles of the $\gamma q \to Wb$ subprocess and, $\hat{s}=z^2s$ with $z=\xi x$. In our paper, we have used Martin, Stirling, Thorne and Watt parton distribution functions \cite{pdf}. During calculations, we have taken the quark virtuality $Q^{'2}=m_t^2$.
In all the results presented in this work, we impose a cut of pseudorapidity $|\eta|<2.5$ for final state particles from subprocess $\gamma q \to Wb$ since central detectors of the ATLAS and CMS have a pseudorapidity $|\eta|$ coverage of $2.5$.

\section{Phenomenological Analysis}

The subprocess $\gamma q \to Wb$ consists of $s$, $t$ and $u$ channel tree-level SM  diagrams. Additionally, there is a
one tree-level Feynman diagram containing anomalous $tq\gamma$ coupling in Fig.\ref{fig2}. The total polarization summed amplitude squared are given in Appendix.
In our calculations, it is assumed that the center of mass energy of the LHC is $14$ TeV.

The total cross sections as a function of $\kappa_{tq\gamma}$ for four acceptance regions  $0.0015< \xi < 0.5$, $0.0015< \xi <0.15$, $0.1< \xi < 0.5$ and $0.015< \xi <0.15$ are presented in Fig. \ref{fig3}. We see from this figure that total cross sections for the $0.0015< \xi < 0.5$ and $0.0015< \xi < 0.15$ are close to each other.
In Fig. \ref{fig4}, we have plotted the SM and total cross sections of $pp \to pWbX$ as functions of the transverse momentum cut ($p_t$ cut or $p_{t, min}$) of the final state particles for $\kappa_{tq\gamma}=0.01$ and two forward detectors acceptance regions: $0.0015< \xi < 0.5$ and $0.1< \xi < 0.5$. Fig. \ref{fig5} same as Fig. \ref{fig4} but for the other acceptances regions: $0.0015< \xi <0.15$ and $0.015< \xi <0.15$.
As seen from these figures, in actual experiments both angular distribution and the $p_t$ cut can be used to improve the sensitivity bounds since contributions of the new physics and the SM are well separated from each other for high $p_t$ cut regions. Moreover, the acceptance region $0.1< \xi < 0.5$ has almost the same features as the other acceptance regions with a high $p_t$ cut. It can be concluded that a high lower bound of the acceptance region mimics an extra $p_t$ cut. Therefore, in this paper we estimate sensitivity of the $pp \to p \gamma p \to p W b X $ process to be $tq\gamma$ anomalous couplings using two different statistical analysis methods.  First, we use
a Poisson distribution, which is the appropriate sensitivity analysis since the number of SM events with these cuts are small enough.  In this statistical analysis, the number of observed events are assumed to be equal to the SM prediction $N_{obs}=S\times E\times BR\times L\times\sigma_{SM}=N_{SM}$. Here S is the survival probability factor, E is the jet reconstruction efficiency and L is the integrated luminosity. We have taken a survival probability factor of $S=0.7$ \cite{khoze1}, and the $b$ jet reconstruction efficiency of $E=0.6$.
We consider W boson decay leptonically, hence here BR is the branching ratio of W boson to leptons. In the second statistical analysis we have used the $\chi^2$ criterion without a systematic error which is given by

\begin{eqnarray}
\chi^2=\left(\frac{\sigma_{SM}^i-\sigma_{NEW}^i}{\sigma_{SM}^i\delta}\right)^2
\end{eqnarray}
where $\sigma_{NEW}$ is the total cross section including $SM$ and new physics and $\delta$ is the statistical error.
We show the sensitivity of the 95\% C.L. parameter $\kappa_{tq\gamma}$ as a function of integrated LHC luminosity for $0.0015< \xi <0.5$ and $0.1< \xi <0.5$ in Fig. $6$ and $0.0015< \xi <0.5$, $0.0015< \xi <0.15$ in Fig.$7$. We set $p_t>30$ GeV and $|\eta|<2.5$ in these figures.

During calculations we considered all tree-level SM contributions for the subprocess $\gamma q \to Wb$
(Fig.\ref{fig2}). These generate major backgrounds. On the other hand,
the leading order background to this process might be coming from the pomeron-quark interaction. A
pomeron emitted from one of the incoming proton beam can collide with the other proton's quarks
and same final state particles can take place. However, when examined in detail it can been seen that this background
process is expected to have a quite small influence on limits of the anomalous coupling. In DIS
process the virtuality of the struck quark is quite high. In this work, we take the
the virtuality of the struck quark $Q^2 = m_t^2$.
Hence, when a pomeron collides with a quark it may be dissociate into partons.
Pomeron remnants can be caught by the calorimeters and this background
can be removed. Morever, the survival probability for a pomeron exchange is
quite smaller than the survival probability of induced photons. Hence, even if the background from
pomeron exchange can not be eliminated, it does not affect the bounds on anomalous coupling \cite{albrow,inanc3}.

In low luminosity values the pile-up of events is negligible in $\gamma p$ interactions at the LHC. However, these backgrounds can be suppressed by using exclusivity conditions, kinematics and timing constraints at high luminosity \cite{pile0, pile1, pile2, albrow}. For these purposes, we give the sensitivity bounds for between luminosity values of $1-200$ fb$^{-1}$ in Figs. \ref{fig6} and \ref{fig7}. As seen from these figures, SM backgrounds could be smaller than $10$ depending on the integrated luminosity. Therefore, in these kinematical regions we have used Poisson analysis for the $N_{SM}<10$ and, we have used $\chi^2$ criterion for $N_{SM}>10$. We understand from the figures that the best sensitivity has been obtained in the $0.0015< \xi <0.5$ case. In Fig. \ref{fig8} we show the 95\% C.L. lower bounds for $\kappa_{tq\gamma}$ as a function of integrated LHC luminosity for $0.0015< \xi <0.5$, $0.1< \xi <0.5$ and $p_t>500$ GeV. Fig. \ref{fig9} same as Fig. \ref{fig8} but for $0.0015< \xi <0.15$ and $0.015< \xi <0.15$. In this high $p_t$ cut region, SM events smaller than $10$ for all of the luminosity values as seen from Figs. \ref{fig4} and \ref{fig5}. Hence, in Figs. \ref{fig8} and \ref{fig9} we use only Poisson analysis. These figures show that the obtained sensitivity bounds in Figs. \ref{fig6} and \ref{fig7} are better than in Figs.\ref{fig8} and \ref{fig9}. However, high $p_t$ cut regions have a very clean environment. Therefore, any signal which conflicts with the SM expectations would be a credible clue for there being something beyond the SM.

\section{Conclusions}

By using very forward detectors, the LHC can be designed as a high energy photon-photon and photon-proton collider. There is no existing high energy photon-photon, photon-proton collider with this property. The process $pp \to p\gamma p \to pWbX$ provides fewer backgrounds than the pure DIS process due to one of the incoming protons being intact after the collision. The detection of the intact protons in forward detectors make it possible to determine the momentum of the quasi-real photons. This situation may be useful in determining the kinematics of the process. Moreover, anomalous $tq\gamma$ couplings might also be uniquely revealed in single top photoproduction \cite{albrow}.

In these motivations, we have analysed the potential of the $pp \to p\gamma p \to pWbX$ at the LHC to probe anomalous $tq\gamma$ couplings for four forward detector acceptances $0.0015< \xi < 0.5$, $0.0015< \xi < 0.15$, $0.015< \xi < 0.15$ and, $0.1< \xi <0.5$. We determined that this photon-induced process has an important potential to examine anomalous $tq\gamma$ couplings. We have investigated the sensitivity bounds for $p_t>30$ GeV and $p_t>500$ GeV regions.
The sensitivity bounds on $tq\gamma$ coupling are better than the current experimental results even at luminosity value of $1$ fb$^{-1}$. For this luminosity value, bounds on $tq\gamma$ coupling can be improved $18$ times with respect to present experimental datas as seen from Fig.\ref{fig6}.

On the other hand, we show that obtained results improve the sensitivity bounds  by up to a factor of $116$ for $0.0015< \xi < 0.5$ with respect to current experimental data as seen from Fig. \ref{fig6}. Furthermore, for $p_t>500$ GeV, the results improve the sensitivity bounds on $tq\gamma$ couplings by up to a factor of $38$ for $0.0015< \xi < 0.5$. These high $p_t$ cut regions can give extra opportunities to search for new physics with very low backgrounds. As a result, forward detectors provide an enhancement of the physics studied at the LHC.

\section{Appendix}

The total polarization summed amplitude squared which
consists of SM, new physics and interference parts has been obtained
in functions of the Mandelstam invariants $\hat{s}$, $\hat{t}$ and $\hat{u}$ as follows,

\begin{eqnarray}
|M_{1}|^{2}=-\frac{g_{e}^{2}g_{w}^{2}V_{bq}^{2}e_{u}^{2}}{m_{w}^{2}s} (m_{w}^{4}-(\hat{s}-\hat{t}+\hat{u})m_{w}^{2}+(\hat{s}-m_{b}^{2})\hat{u}),
\end{eqnarray}

\begin{eqnarray}
|M_{2}|^{2}=&&\frac{g_{e}^{2}g_{w}^{2}V_{bq}^{2}e_{b}^{2}}{m_{w}^{2}(\hat{t}-m_{b}^{2})^{2}}(-2\hat{t}m_{b}^{4}+(5m_{w}^{4}+(\hat{s}-5\hat{t}-\hat{u})m_{w}^{2}+
\hat{t}(2\hat{s}+3\hat{u}))m_{b}^{2}+ \nonumber \\ &&\hat{t}(-m_{w}^{4}+(-\hat{s}+\hat{t}+\hat{u})m_{w}^{2}-\hat{t}\hat{u})), \nonumber \\
\end{eqnarray}

\begin{eqnarray}
|M_{3}|^{2}=&&\frac{g_{e}^{2}g_{w}^{2}V_{bq}^{2}}{4m_{w}^{4}(\hat{u}-m_{w}^{2})^{2}} ((m_{w}^{2}-\hat{u})(3m_{w}^{2}-\hat{u})(m_{b}^{2}+m_{w}^{2}-\hat{t}-\hat{u})(\hat{u}-\hat{t})+ \nonumber \\
&&s(-3m_{w}^{6}+(3\hat{s}-14\hat{t}+4\hat{u})m_{w}^{4}-\hat{u}(4(\hat{s}+\hat{t})+\hat{u})m_{w}^{2}+(\hat{s}+2\hat{t})\hat{u}^{2}+ \nonumber \\
&&m_{b}^{2}(11m_{w}^{2}-3\hat{u})(m_{w}^{2}+\hat{u}))), \nonumber \\
\end{eqnarray}

\begin{eqnarray}
|M_{4}|^{2}=&&\frac{g_{e}^{2}g_{w}^{2}\kappa_{tq\gamma}^{2}e_{t}^{2}V_{tb}^{2}}{m_{w}^{2}\Lambda^{2}((\hat{s}-m_{t}^{2})^{2}+\Gamma_{t}^{2}m_{t}^{2})}\hat{s}((-m_{w}^{4}+
(\hat{s}-\hat{t}+\hat{u})m_{w}^{2}+ \nonumber \\ &&(m_{b}^{2}-\hat{s})\hat{u})m_{t}^{2}+\hat{s}(-m_{w}^{4}+(\hat{s}+\hat{t}-\hat{u})m_{w}^{2}+(m_{b}^{2}-\hat{s})\hat{t})), \nonumber \\
\end{eqnarray}

\begin{eqnarray}
2Re(M_{1}^{\dag}M_{2})=&&\frac{g_{e}^{2}g_{w}^{2}V_{bq}^{2}e_{u}e_{b}}{m_{w}^{2}\hat{s}(\hat{t}-m_{b}^{2})}((\hat{s}-3\hat{t}-\hat{u})m_{b}^{4}+
(4m_{w}^{4}+(\hat{s}-\hat{t}+\hat{u})m_{w}^{2}-2\hat{s}^{2}+
\nonumber \\ &&\hat{u}^{2}+4\hat{s}\hat{t}+\hat{s}\hat{u}+3\hat{t}\hat{u})m_{b}^{2}-(m_{w}^{2}-\hat{s}-\hat{t})(\hat{s}-\hat{t})^{2}-(\hat{s}+\hat{t})\hat{u}^{2}+ \nonumber \\
&&(-4m_{w}^{4}+(\hat{s}+\hat{t})m_{w}^{2}-2\hat{s}\hat{t})\hat{u}), \nonumber \\
\end{eqnarray}

\begin{eqnarray}
2Re(M_{1}^{\dag}M_{3})=&&\frac{g_{e}^{2}g_{w}^{2}V_{bq}^{2}e_{u}}{4m_{w}^{4}\hat{s}(m_{w}^{2}-\hat{u})}(2(\hat{s}+4\hat{t}-4\hat{u})m_{w}^{6}+
(\hat{s}^{2}+13\hat{t}\hat{s}-7\hat{u}\hat{s}-12\hat{t}^{2}+12\hat{u}^{2})m_{w}^{4}+
\nonumber \\ &&(-3\hat{s}^{3}+(\hat{u}-6\hat{t})\hat{s}^{2}+(-3\hat{t}^{2}-3\hat{u}\hat{t}+2\hat{u}^{2})\hat{s}+4(\hat{t}-\hat{u})(\hat{t}+\hat{u})^{2})m_{w}^{2}+ \nonumber \\
&&\hat{s}\hat{u}(\hat{s}+\hat{t}-\hat{u})(\hat{s}+\hat{t}+\hat{u})+m_{b}^{2}(-4(\hat{s}-2\hat{t}+2\hat{u})m_{w}^{4}+(7\hat{s}^{2}+5\hat{t}\hat{s}-3\hat{u}\hat{s} \nonumber \\
&&-4\hat{t}^{2}+4\hat{u}^{2})m_{w}^{2}+\hat{s}\hat{u}(-\hat{s}-3\hat{t}+\hat{u}))), \nonumber \\
\end{eqnarray}

\begin{eqnarray}
2Re(M_{1}^{\dag}M_{4})=&&\frac{m_{t}e_{u}g_{e}^{2} g_{w}^{2}\kappa_{tq\gamma} e_{t}V_{bq}V_{tb}(m_{t}^{2}-\hat{s})}{\Lambda m_{w}^{2}((\hat{s}-m_{t}^{2})^{2}+\Gamma_{t}^{2}m_{t}^{2})}
(-3m_{w}^{4}+(3\hat{s}-\hat{t}+\hat{u})m_{w}^{2}+(m_{b}^{2}-\hat{s})(\hat{t}+2\hat{u})), \nonumber \\
\end{eqnarray}

\begin{eqnarray}
2Re(M_{2}^{\dag}M_{3})=&&\frac{g_{e}^{2}g_{w}^{2}V_{bq}^{2}e_{b}}{4m_{w}^{4}(\hat{u}-m_{w}^{2})(\hat{t}-m_{b}^{2})}
(2(4\hat{s}+\hat{t}-4\hat{u})m_{w}^{6}+ \nonumber \\
&&(-12\hat{s}^{2}+13\hat{t}\hat{s}+(\hat{t}-4\hat{u})(\hat{t}-3\hat{u}))m_{w}^{4}+(4\hat{s}^{3}+(4\hat{u}-3\hat{t})\hat{s}^{2}-(6\hat{t}^{2}+ \nonumber \\
&&3ut+4u^{2})s-3t^{3}-4u^{3}+2tu^{2}+t^{2}u)m_{w}^{2}+tu(s+t-u)(s+t+u)+ \nonumber \\
&&m_{b}^{2}(6m_{w}^{6}-(21\hat{s}+\hat{t}+9\hat{u})m_{w}^{4}+(-5\hat{s}^{2}+13\hat{t}\hat{s}+3\hat{u}\hat{s}+8\hat{t}^{2}+6\hat{u}^{2})m_{w}^{2}+ \nonumber \\
&&\hat{u}(-\hat{s}^{2}-3\hat{t}\hat{s}-4\hat{t}^{2}+\hat{u}^{2}+\hat{t}\hat{u}))+m_{b}^{4}(4m_{w}^{4}-(3\hat{s}+5(\hat{t}+\hat{u}))m_{w}^{2}+ \nonumber \\
&&(\hat{s}+3\hat{t}-\hat{u})\hat{u})) \nonumber \\
\end{eqnarray}

\begin{eqnarray}
2Re(M_{2}^{\dag}M_{4})=&&\frac{m_{t}e_{b}g_{e}^{2}g_{w}^{2} \kappa_{tq\gamma} e_{t}V_{bq}V_{tb}}{\Lambda m_{w}^{2}(\hat{t}-m_{b}^{2})((\hat{s}-m_{t}^{2})^{2}+\Gamma_{t}^{2}m_{t}^{2})}
((m_{t}^{2}-\hat{s})((\hat{s}-\hat{u})m_{b}^{4}+ \nonumber \\
&&(m_{w}^{4}+(3\hat{s}+\hat{t}-\hat{u})m_{w}^{2}-2\hat{s}^{2}+\hat{u}^{2}+\hat{s}\hat{u})m_{b}^{2}+(m_{w}^{2}-\hat{s})\hat{u}^{2}+ \nonumber \\
&&(m_{w}^{2}-\hat{s})^{2}s-m_{w}^{2}(m_{w}^{2}-\hat{s}+\hat{t})\hat{u})+4\Gamma_{t}m_{t}(m_{b}^{2}+m_{w}^{2}-\hat{s}) \epsilon^{ p_{1} p_{2} p_{3} p_{4}}), \nonumber \\
\end{eqnarray}

\begin{eqnarray}
2Re(M_{3}^{\dag}M_{4})=&&\frac{m_{t}g_{e}^{2}g_{w}^{2} \kappa_{tq\gamma} e_{t}V_{bq}V_{tb}}{2\Lambda m_{w}^{2}(\hat{u}-m_{w}^{2})((\hat{s}-m_{t}^{2})^{2}+\Gamma_{t}^{2}m_{t}^{2})}
((m_{t}^{2}-\hat{s})\hat{s}(-m_{w}^{4}+ \nonumber \\
&&(\hat{s}-8\hat{t}+\hat{u})m_{w}^{2}-\hat{s}\hat{u}+m_{b}^{2}(7m_{w}^{2}+\hat{u}))+(m_{w}^{2}-\hat{u}) \nonumber \\
&&(4\Gamma_{t}m_{t} \epsilon^{ p_{1} p_{2} p_{3} p_{4}}-(m_{t}^{2}-\hat{s})(m_{b}^{2}+m_{w}^{2}-\hat{t}-\hat{u})(\hat{t}-\hat{u}))), \nonumber
\\
\end{eqnarray}

\noindent where $g_{e}$ and $g_w$ are the electromagnetic and weak coupling constants, $m_b$ is the $b$ quark mass and $m_w$ is the W boson mass. $p_1$, $p_2$, $p_3$ and $p_4$ are the momentums of the photon, incoming quark, $W$ boson and $b$ quark, respectively. $V_{bq}$ and $V_{tb}$ are the corresponding CKM matrix elements. $e_{u}$($e_{b}$) is the electric charge of the $u$($b$) quark. Also, $\Gamma_{t}$ is the total decay width of the top quark. We have neglected the mass of the incoming quarks.

\pagebreak

\pagebreak

\begin{figure}
\includegraphics{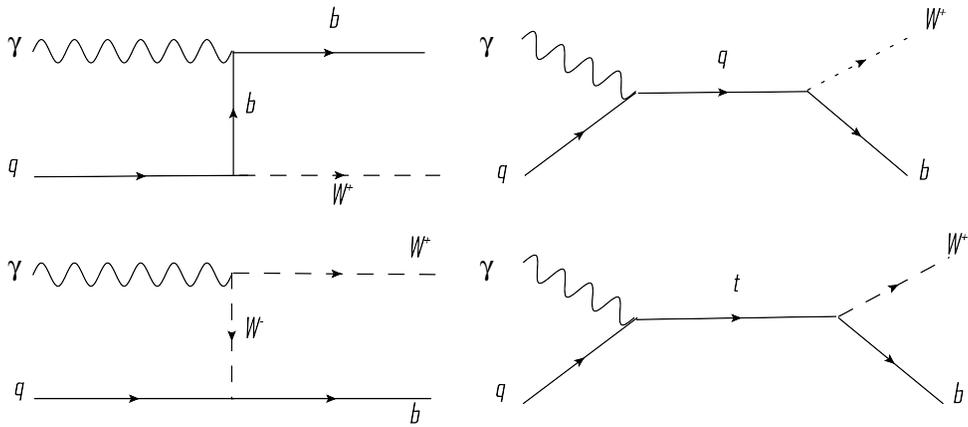}
\caption{Tree level Feynman diagrams for the subprocess $\gamma q \to Wb$ ($q=u,c$) in the presence of the anomalous $tq \gamma$ couplings.
\label{fig2}}
\end{figure}

\begin{figure}
\includegraphics{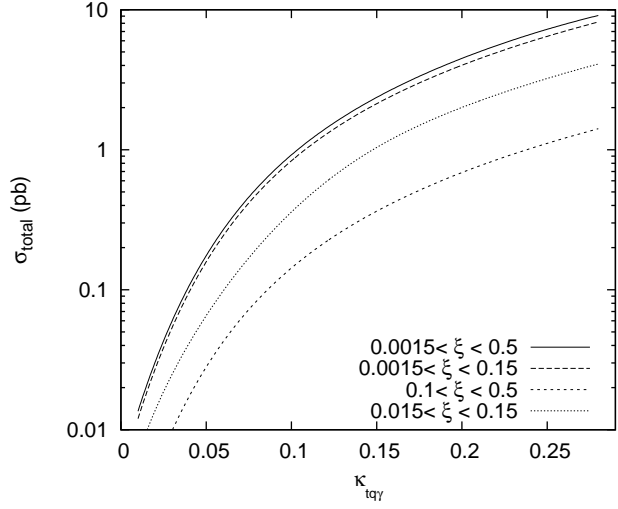}
\caption{The total cross sections of $pp \to p\gamma p \to pWbX$ as a function of anomalous $tq \gamma$ coupling ($\kappa_{tq\gamma}$) for four different forward detector acceptances stated in the figure. It is assumed that the center of mass energy of the LHC is $14$ TeV. Also, we impose cuts
$|\eta|<2.5$ and $p_t >30$ GeV.
\label{fig3}}
\end{figure}

\begin{figure}
\includegraphics{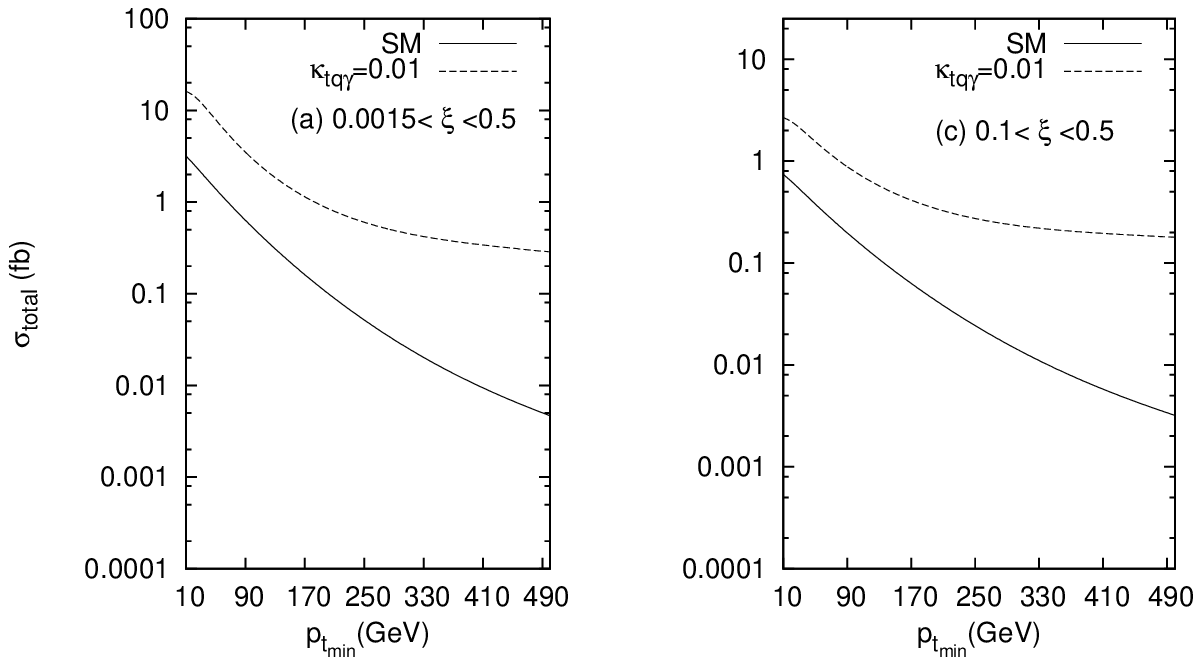}
\caption{Cross sections of $pp \to p\gamma p \to pWbX$  as a function of
the transverse momentum cut on the final state particles for two forward detector acceptances: $0.0015< \xi <0.5$ and $0.1< \xi <0.5$. Solid lines are for the SM and the dotted lines are for the total cross sections with $\kappa_{tq\gamma}=0.01$.
We impose cuts $|\eta|<2.5$ and $p_t >30$ GeV.
\label{fig4}}
\end{figure}

\begin{figure}
\includegraphics{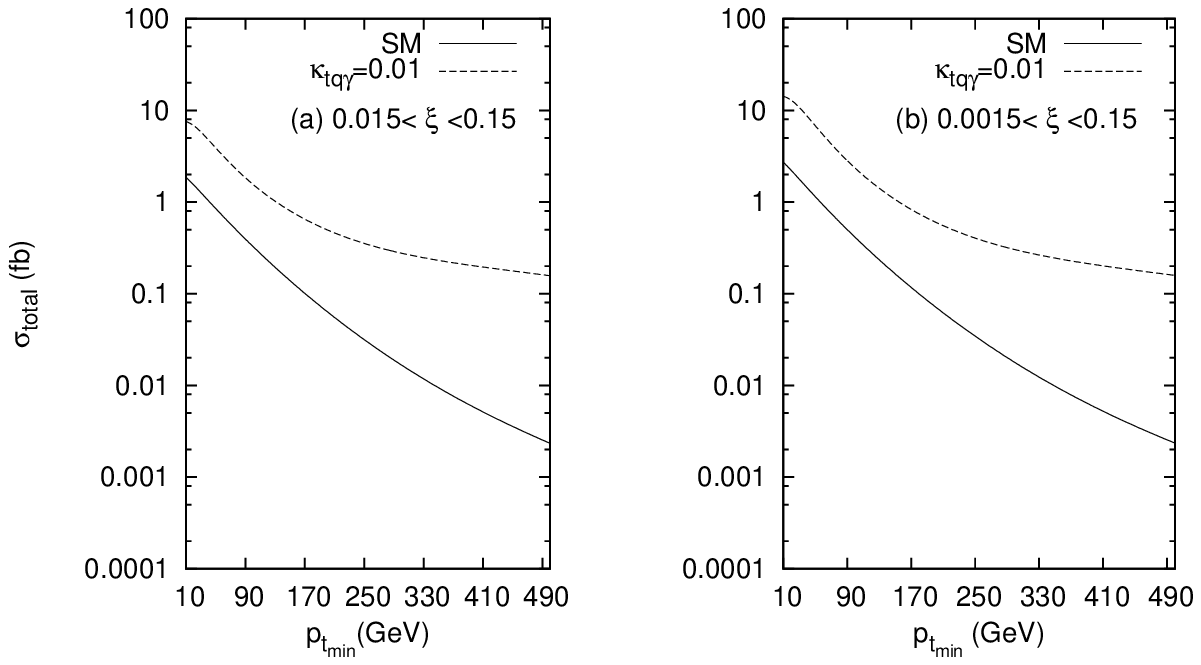}
\caption{Cross sections of $pp \to p\gamma p \to pWbX$  as a function of
the transverse momentum cut on the final state particles for two forward detector acceptances: $0.015< \xi <0.15$ and $0.0015< \xi <0.15$. Solid lines are for the SM and the dotted lines are for the total cross sections with $\kappa_{tq\gamma}=0.01$.
We impose cuts $|\eta|<2.5$ and $p_t >30$ GeV.
\label{fig5}}
\end{figure}

\begin{figure}
\includegraphics{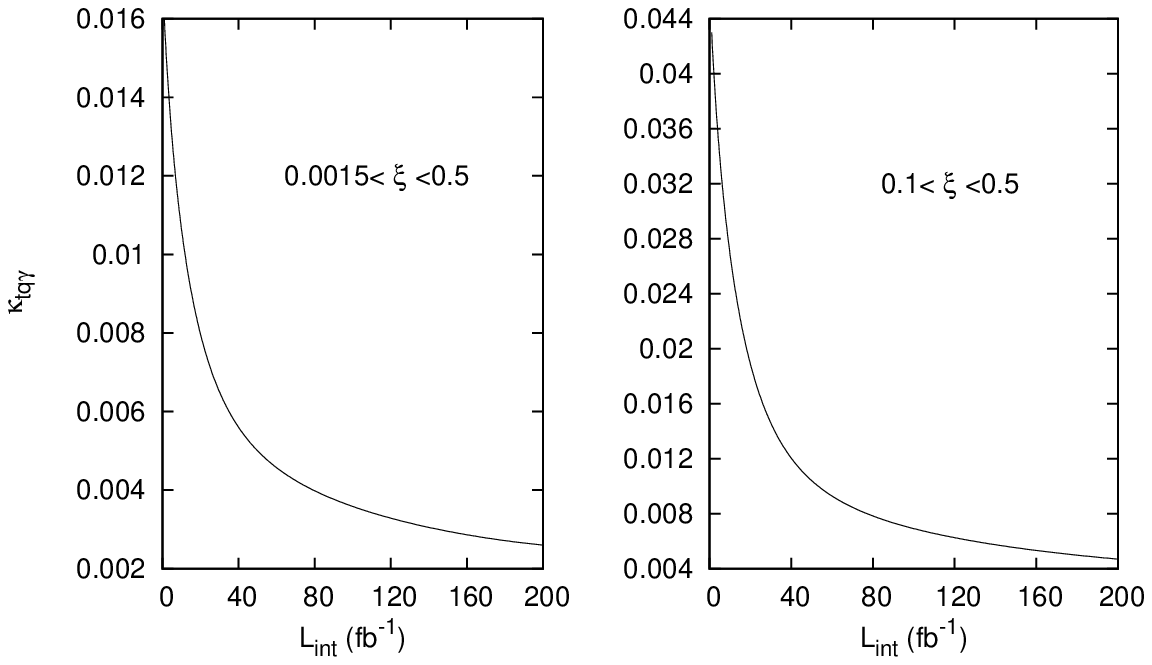}
\caption{$95\%$ C.L. sensitivity bounds for $\kappa_{tq\gamma}$ as a function of integrated LHC luminosity for two forward detector acceptances: $0.0015< \xi <0.5$ and $0.1< \xi <0.5$. We impose the following cuts: $p_t >30$ GeV and $|\eta|<2.5$.
\label{fig6}}
\end{figure}

\begin{figure}
\includegraphics{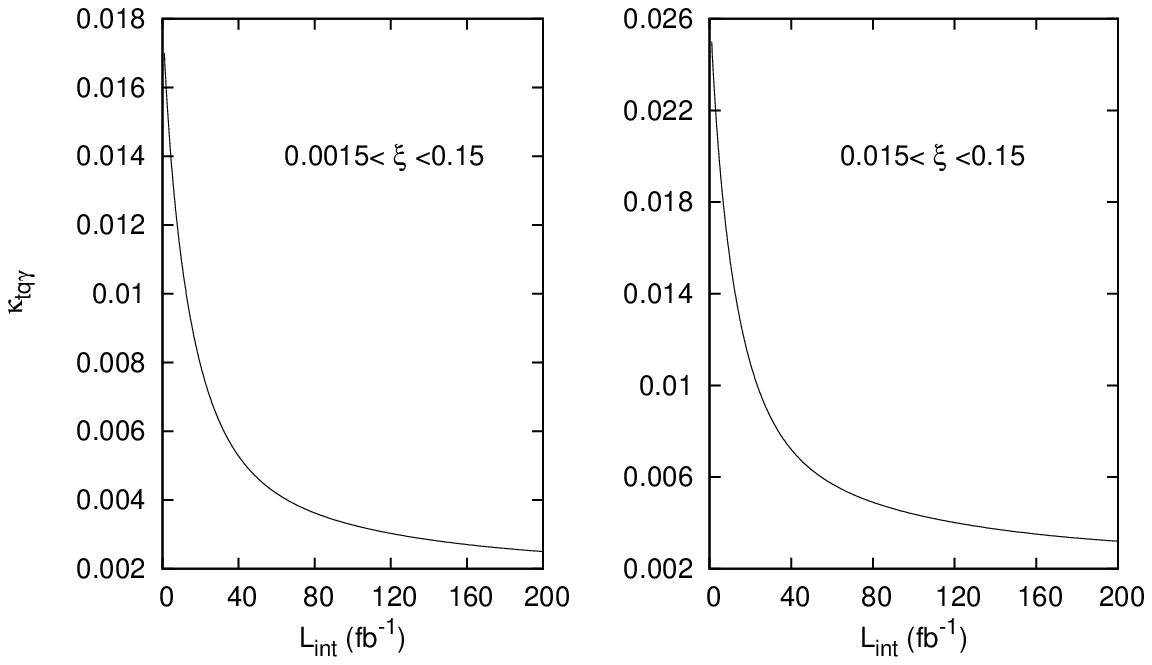}
\caption{$95\%$ C.L. sensitivity bounds for $\kappa_{tq\gamma}$ as a function of integrated LHC luminosity for two forward detector acceptances: $0.0015< \xi <0.15$ and $0.015< \xi <0.15$. We impose the following cuts: $p_t >30$ GeV and $|\eta|<2.5$.
\label{fig7}}
\end{figure}

\begin{figure}
\includegraphics{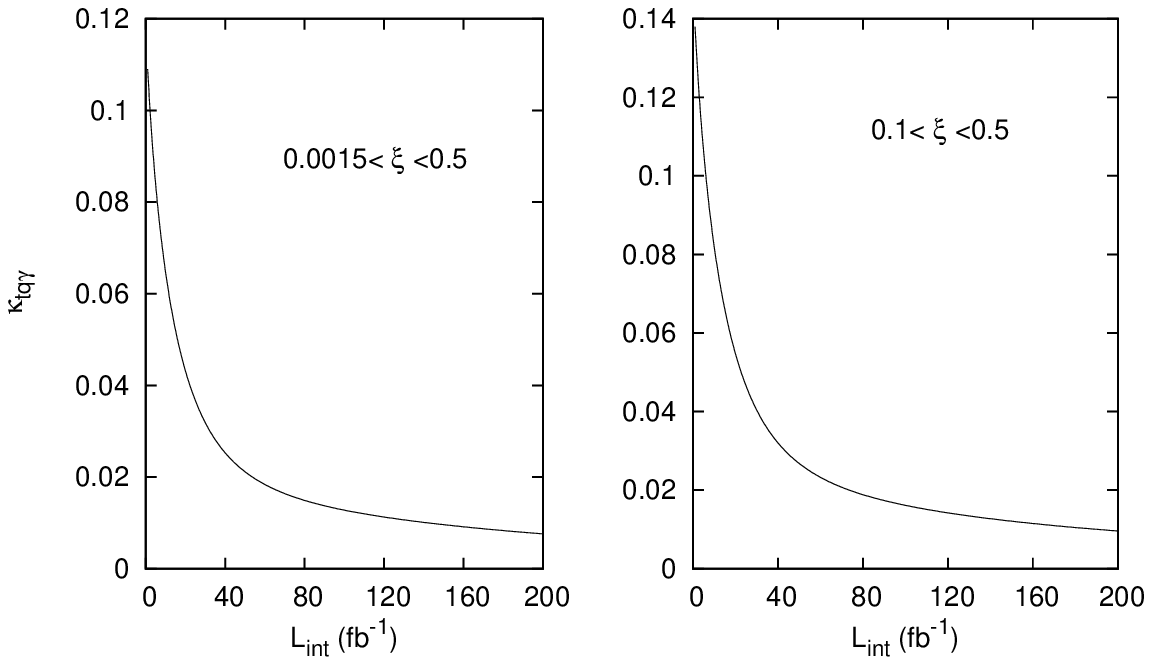}
\caption{$95\%$ C.L. sensitivity bounds for $\kappa_{tq\gamma}$ as a function of integrated LHC luminosity for two forward detector acceptances: $0.0015< \xi <0.5$, and $0.1< \xi <0.5$. We impose the following cuts: $p_t >500$ GeV and $|\eta|<2.5$.
\label{fig8}}
\end{figure}

\begin{figure}
\includegraphics{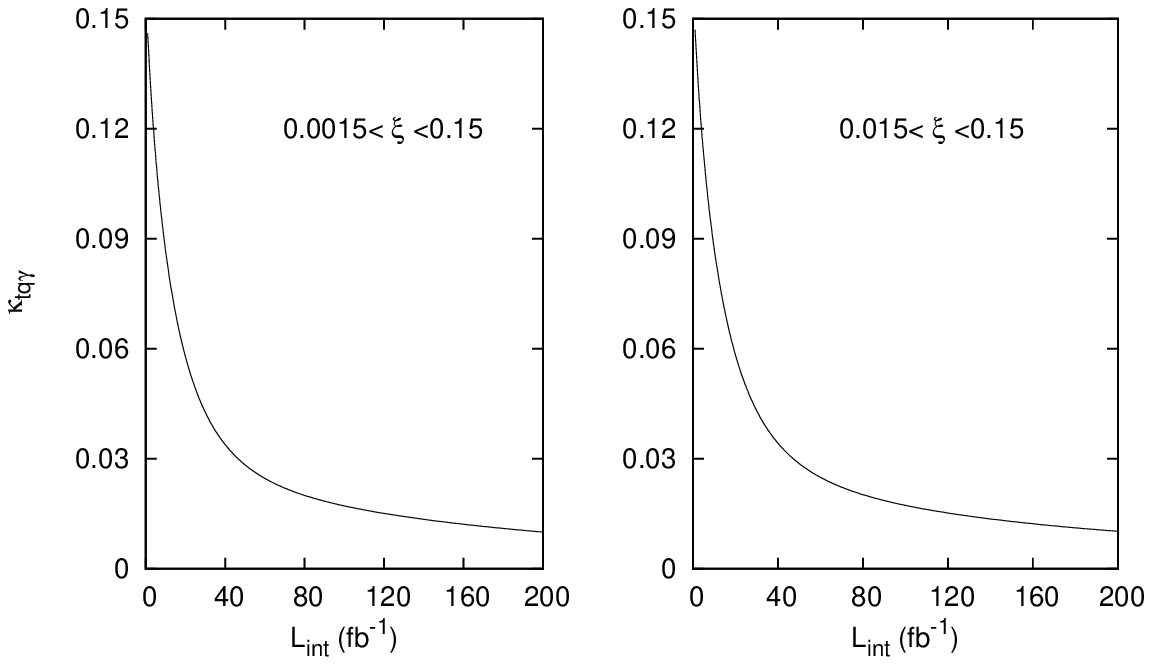}
\caption{$95\%$ C.L. sensitivity bounds for $\kappa_{tq\gamma}$ as a function of integrated LHC luminosity for two forward detector acceptances: $0.0015< \xi <0.15$ and $0.015< \xi <0.15$. We impose the following cuts: $p_t >500$ GeV and $|\eta|<2.5$.
\label{fig9}}
\end{figure}

\end{document}